\documentclass[aps,prl,twocolumn,groupedaddress,nofootinbib]{revtex4-1}
\usepackage{amsmath}
\usepackage{graphicx}
\usepackage{sistyle}
\usepackage{gensymb}
\usepackage{bm}
\usepackage{natbib}
\usepackage{dsfont}
\usepackage{epsfig}

\makeatletter
\newcommand\footnoteref[1]{\protected@xdef\@thefnmark{\ref{#1}}\@footnotemark}
\makeatother

\usepackage{color}

\begin{document}

\title{Force and acceleration sensing with optically levitated nanogram masses at microkelvin temperatures}

\author{Fernando Monteiro} 
\email{fernando.monteiro@yale.edu}
\author{Wenqiang Li} \altaffiliation{Also at College of Optical Science and Engineering, Zhejiang University, Hangzhou, China}
\author{Gadi Afek}
\author{Chang-ling Li} \altaffiliation{Also at Ecole CentraleSup\'elec, Gif-sur-Yvette and Universit\'e Paris-Sud, Orsay, France}
\author{Michael Mossman} 
\author{David C. Moore}
\affiliation{Wright Laboratory, Department of Physics, Yale University, New Haven, CT 06520, USA}

\begin{abstract}

This paper demonstrates cooling of the center-of-mass motion of 10~$\mu$m-diameter optically levitated silica spheres to an effective temperature of $50\pm22~\mu$K, achieved by minimizing the technical pointing noise of the trapping laser. This low noise leads to an acceleration and force sensitivity of $95\pm41$ n$g/\sqrt{\mathrm{Hz}}$ ($g = 9.8$ m/s$^2$) and $0.95\pm0.11$~aN$/\sqrt{\mathrm{Hz}}$, respectively, at frequencies near 50~Hz. This force sensitivity is comparable to that demonstrated for optically levitated nanospheres that are $10^4$ times less massive, corresponding to an acceleration sensitivity that is several orders of magnitude better. It is further shown that under these conditions the spheres remain stably trapped at pressures of $\sim 10^{-7}$~mbar with no active cooling for periods longer than a day. Feedback cooling is still necessary in the moderate-pressure regime, motivating a comprehensive study of the loss mechanisms of the microspheres and providing better understanding of the requirements for feedback-free optical trapping in vacuum. This work can enable high-sensitivity searches for accelerations and forces acting on micron-sized masses, including those that could be produced by new physics beyond the Standard Model.

\end{abstract}

\maketitle

\section{Introduction}

Following the pioneering work of Ashkin and Dziedzic~\cite{Ashkin:1971,Ashkin:1976, Ashkin:1977}, optical trapping of massive objects in air and vacuum has led to the development of precise force and acceleration sensors at the micro- and nano-scale~\cite{gieseler:2013,Ranjit:2016,Hempston:2017,Novotny_static:2018,Ranjit:2015,Rider:2017,Blakemore_3D_microscope:2019,Acceleration_2017}. The high-sensitivity provided by optically levitated microspheres in vacuum has enabled applications including new searches for physics beyond the Standard Model~\cite{Moore:2014,Rider:2016}. Recent developments using optically trapped nanospheres allow for near ground state~\cite{Novotny2019,Novotny:2020} and ground state~\cite{Aspelmeyer:2019} cooling of such particles, possibly enabling tests of quantum mechanics at mesoscopic scales.

Force and acceleration sensors that employ levitated masses in high vacuum can be decoupled from most sources of environmental noise, such as vibrations and collisions with residual gas molecules. Such sensors can be implemented by the use of optical~\cite{Ashkin:1971,gieseler:2013,Ranjit:2015,Ranjit:2016,Rider:2017,Acceleration_2017,Hempston:2017,Novotny_static:2018,Blakemore_3D_microscope:2019} or magnetic levitation of microsized objects~\cite{DUrso:2018,Ulbricht:2019,2019arXiv191210397G}. While recent implementations of magnetic levitation of microspheres predict attainable sensitivities that surpass current implementations achieved with optically levitated objects~\cite{DUrso:2018,Ulbricht:2019}, optical trapping provides technical advantages that can favor its use in certain applications of force and acceleration sensing. In particular, optical trapping avoids the need for large magnetic field gradients or superconducting objects, allowing simple, room-temperature force sensors and accelerometers. It also permits the use of low numerical aperture (NA) beams~\cite{Ashkin:1971,Moore:2014,Rider:2016,Acceleration_2017, Rotation_paper_2018}, enabling probes to be positioned in close proximity to the object. Such access is essential for searches for new interactions that could depend exponentially on separation distance~\cite{Nelson:2003, Geraci:2008,Rider:2016, Deca:2016, Yoshioka:2018}, and for the control of stray electric and magnetic fields that can couple to levitated objects and induce unwanted background forces.

It has been recently shown that technical pointing noise of the trapping laser limits the acceleration sensitivity obtained with optically trapped microspheres in vacuum in previous work~\cite{Ranjit:2016,Acceleration_2017}, rather than laser shot noise or residual gas damping. The results presented here include a substantial reduction of this noise, relative to previous demonstrations, consequently allowing for an order-of-magnitude improvement of the state-of-the-art acceleration sensitivity measured with optically levitated objects. Such improvement also allows for a force sensitivity comparable to that of nanopheres that are $10^4$ less massive~\cite{gieseler:2013,Ranjit:2016,Hempston:2017,Novotny_static:2018}, directly translating into an acceleration sensitivity that is more than four orders of magnitude lower. Together with active feedback cooling, the lower noise levels allow an effective center of mass (COM) temperature to be reached that is more than an order of magnitude lower than previously achieved for micron-sized spheres in an optical levitation setup~\cite{Li:2011,Ranjit:2015}.

\begin{figure*}[ht!]
    \centering
    \includegraphics[width=\textwidth]{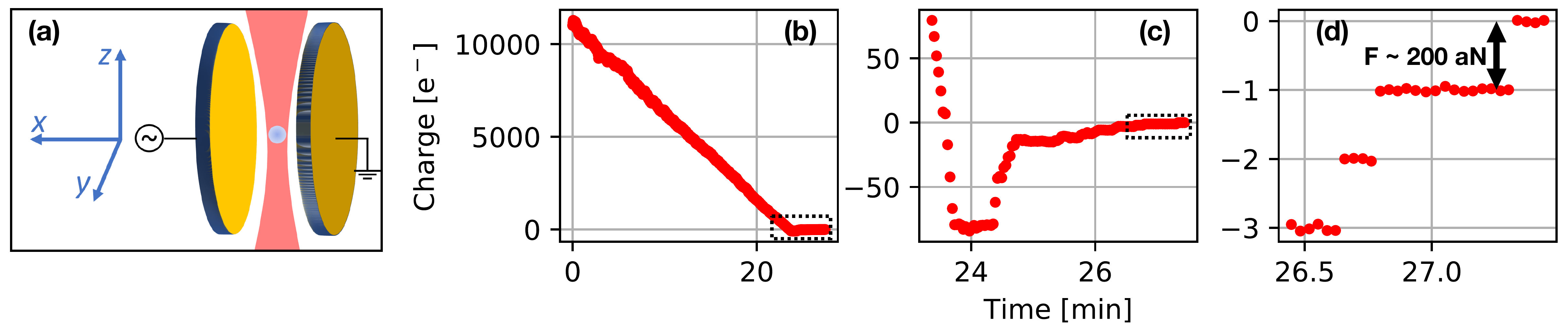}
    \caption{(a) Schematic of the setup used to discharge the sphere and the corresponding coordinate system definition. (b) Net number of electrons present on the sphere as a function of time as the sphere is discharged, starting from an initial charge upon loading of magnitude $>10^4\ e$. (c) Zoom in on the region highlighted by the dashed line in the previous plot, which includes an inversion in the polarity of the net charge. (d) Further zoom in on the single $e$ steps observed in the region highlighted by the dashed line in the previous plot. For plots (b)-(d), the charging rate varies due to changes in the electric field and the rate of flashes from the UV lamp, as described in the text. Each data point corresponds to an integration time of 1~s with an applied electric field at $\sim$36~Hz.}
    \label{charge}
\end{figure*}

Ashkin and Dziedzic's original demonstration of optical trapping in high vacuum was performed without feedback to damp the sphere's motion by employing low optical absorption materials including high-purity fused quartz and silicone oil droplets to minimize heating effects and radiometric forces~\cite{Ashkin:1976}. This low absorption allowed the spheres to be pumped through moderate vacuum pressures and trapped for periods up to $\sim$30 minutes at high vacuum without the use of feedback cooling~\cite{Ashkin:1976}. The present work reproduces---for the first time following the pioneering demonstration---the ability to optically trap microspheres in high vacuum without the use of feedback cooling. Stable, feedback-free levitation is demonstrated even with the use of microspheres with substantially higher optical absorption than used in Ref.~\cite{Ashkin:1976}, and for time periods exceeding one day.

Feedback-free optical levitation in vacuum could substantially simplify modern implementations of force sensors and accelerometers based on optically levitated micron-sized masses in high vacuum. However, feedback cooling is still necessary to prevent sphere loss during the transition between low vacuum (where the trap is initially loaded) to high vacuum. This motivates the study of the trap dynamics and loss mechanisms, including the effects of the internal temperature gradients induced by the laser at these pressures. Future work to minimize these effects or permit loading of microspheres into the trap directly in the high vacuum environment~\cite{Northup:2019} could enable a new class of simple implementations for optical traps in vacuum.

\section{Experimental setup}

The experimental setup, schematically shown in Fig.~\ref{charge}~(a), is similar to that detailed in~\cite{Acceleration_2017}. A vertically oriented 1064~nm laser is used to levitate the sphere, and two additional 532~nm vertically and horizontally oriented beams are used to image its motion. The trapping beam has a low NA $\sim 0.018$, allowing a long working distance (75~mm) between the focusing optics and trapping location. This long working distance allows for two 25~mm diameter electrodes aligned in a parallel plate configuration to be positioned around the trapping region, in order to apply a uniform electric field in the vicinity of the sphere~\cite{Acceleration_2017}. These electrodes are mounted on a vacuum-compatible translation stage so that their separation is controllable between $\sim$1--20~mm. For the measurements reported here, the electrodes are positioned $3.3 \pm 0.4$~mm apart.

The imaging beams are split by three sharp-edged mirrors onto balanced photodiodes to measure all three degrees of freedom (DOF) of the sphere's COM motion. The horizontal imaging beam is used to measure the COM motion in the vertical degree of freedom (denoted here as the $z$ coordinate) while the vertical imaging beam is used to measure the motion in the radial directions ($x$ and $y$). The signals recorded by these ``in-loop'' sensors are used as inputs to a feedback loop that modulates the displacement and amplitude of the trapping beam to provide feedback cooling (see, e.g.~\cite{Ashkin:1977,Li:2011,Moore:2014, Acceleration_2017}), which is implemented following the same general procedure as~\cite{Acceleration_2017}. In addition, an identical ``out-of-loop'' balanced photodiode is positioned in the horizontal imaging beam to record the motion with a beam and sensor that are fully independent from those used in the feedback loop. This out-of-loop sensor allows the COM motion in the $x$ direction to be measured while avoiding unwanted noise cancellation effects (i.e., ``noise squashing'')~\cite{Rugar:2007,Quidant:2019,Novotny2019}. To reduce trapping laser pointing noise, all beams are coupled into single mode optical fibers prior to entering a sealed enclosure holding all free space optics encountered by the beam prior to the optical trap. This enclosure isolates the fiber launch and supporting optics from acoustic noise and air currents~\cite{kwee:2007,Kwee:2008}, and can be pumped to pressures $\lesssim 0.1$~mbar.

The commercially produced amorphous SiO$_2$ spheres are identical to those used in previous work~\cite{Acceleration_2017,Rotation_paper_2018}, and have an average diameter of $10.3\pm1.4$~$\mu$m. The density specified by the sphere manufacturer\footnote{\label{note1}http://www.microspheres-nanospheres.com} is $\rho = 1.8$~g/cm$^3$.

\section{Force calibration}
In this and the following sections, all measurements refer to motion in the $x$ direction as depicted in Fig.~\ref{charge}~(a). While similar results could be obtained for motion in the $y$ direction, the electrodes enable an accurate calibration of the voltage output of the balanced photodiode to a known applied electric force on the sphere, providing a corresponding displacement relative to the equilibrium position of the trap. This calibration is performed following the same procedure as~\cite{Moore:2014,Acceleration_2017}. The voltage measured by the photodiode at time $t$, $V(t)$, is converted to the corresponding position of a sphere with charge $Q$ measured relative to the equilibrium position, $x(t)$, using $x(t) = \frac{QE_0}{M\omega Z_0(\omega)}\sin{(\omega t + \phi)}$, where $\omega$ is the angular frequency of the electric field $\mathbf{E} = E_0 \sin{(\omega t)} \mathbf{\hat{x}}$ with amplitude $E_0$, $M$ is the mass of the sphere, $Z_0(\omega) = \sqrt{(\omega_{0}^{2} - \omega^2)^2 + \omega^2\Gamma^2} $ is the magnitude of the impedance of the harmonic oscillator at angular frequency $\omega$, and $\phi$ is the relative phase of the sphere's motion with respect to the electric field. The damping factor $\Gamma$ and the resonance frequency $\omega_0$ are obtained by fitting the power spectrum of the sphere's motion for this DOF. Using the known electric charge of the sphere, typically $Q = -e$ for the measurements described here, a calibration factor $x(t)/V(t)$ can be determined.

A known net electric charge of a single $e$ on the sphere can be obtained by a controlled charging process employing ultra-violet (UV) light~\cite{Ashkin1980}. Electrons are added to (or removed from) the levitated sphere by shining light from a xenon flash lamp on the sphere's location and its surroundings, while an oscillating electric field is generated by the electrodes. The correlation between the sphere's displacement and the oscillating field is shown in Fig.~\ref{charge}~(b-d),  normalized by the square of the field intensity  and hence proportional to the total charge on the sphere. Figure~\ref{charge}~(d) is used to further transform this correlation into number of electrons present on the sphere. The discrete steps in the response of the sphere's motion serve as an indicator of the total charge that is present in the sphere~\cite{Ashkin1980,Moore:2014,PhysRevA.95.061801,Quidant:2019}. To remove electrons, an electric field amplitude $\gtrsim 50$~V/mm is applied while flashing UV light to ensure electrons ejected from the sphere are drifted away to the nearby electrodes. To avoid pushing the sphere out of the trap during this process, large electric fields of this type are applied at an angular frequency $>10\omega_0$. To add electrons to the sphere, the field amplitude is reduced (typically $\lesssim 2$~V/mm), so that electrons produced by photoelectric ejection on the surrounding gold electrodes can be captured by the sphere. The frequency of the electric field is lowered to $\omega \lesssim \omega_0$ to improve the signal to noise when measuring single electric charge steps.

Following the calibration described above, the sphere is neutralized to have exactly the same number of electrons and protons, $Q = 0$. The measurements of COM temperature and acceleration sensitivity described below were performed using electrically neutral spheres to minimize background forces arising from stray electric fields in the vacuum chamber. Once neutralized, levitated spheres in our system have been observed to remain in the same charge state for several weeks.

\section{Effective temperature}

\begin{figure}
    \centering
    \includegraphics[scale=0.85]{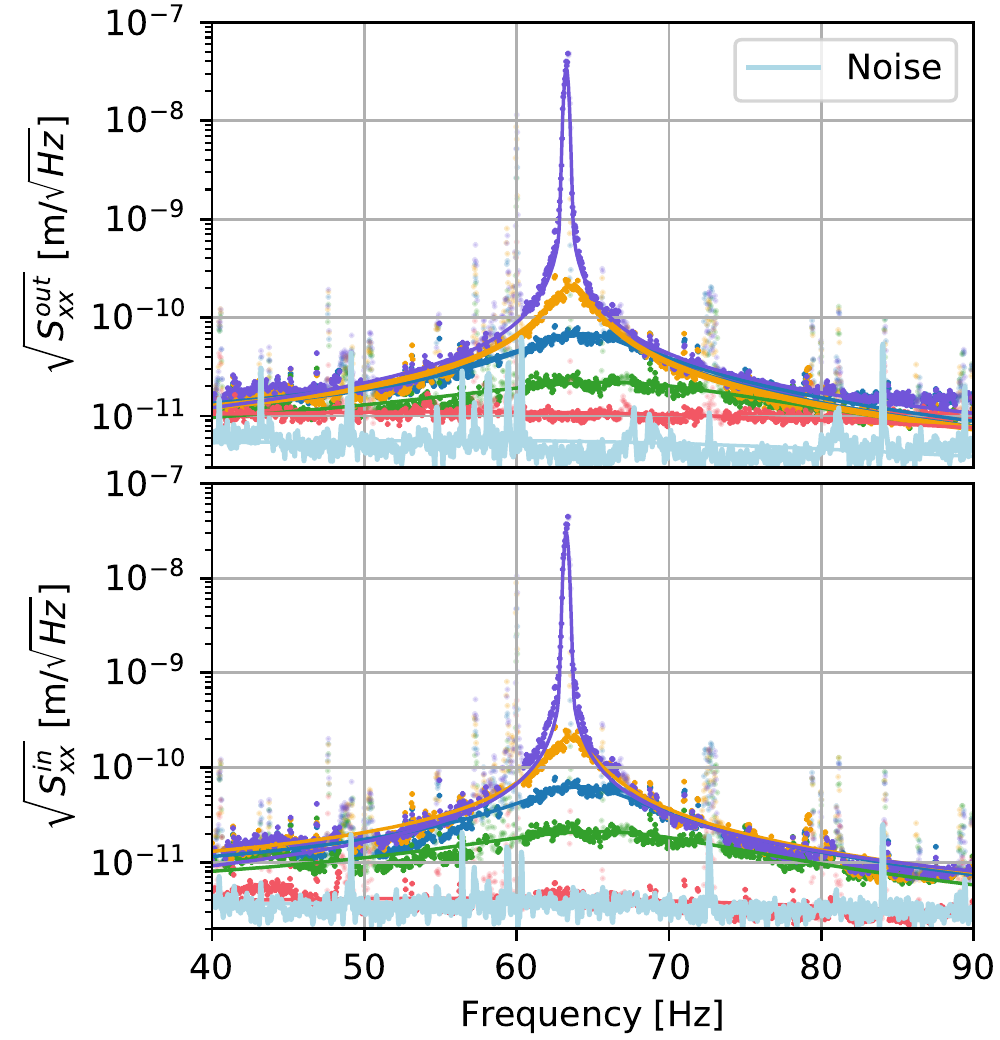}
    \caption{Amplitude spectrum of the $x$ DOF measured by the out-of-loop (upper plot) and in-loop (lower plot) sensors as the feedback gain is varied. The minimum temperature achievable is limited by the noise of the in-loop sensor (lower-most curves), measured without a sphere in the trap. Noise lines that were excluded during data analysis are plotted with lighter colors.}
    \label{psd}
\end{figure}

\begin{figure}
    \centering
    \includegraphics[scale=0.85]{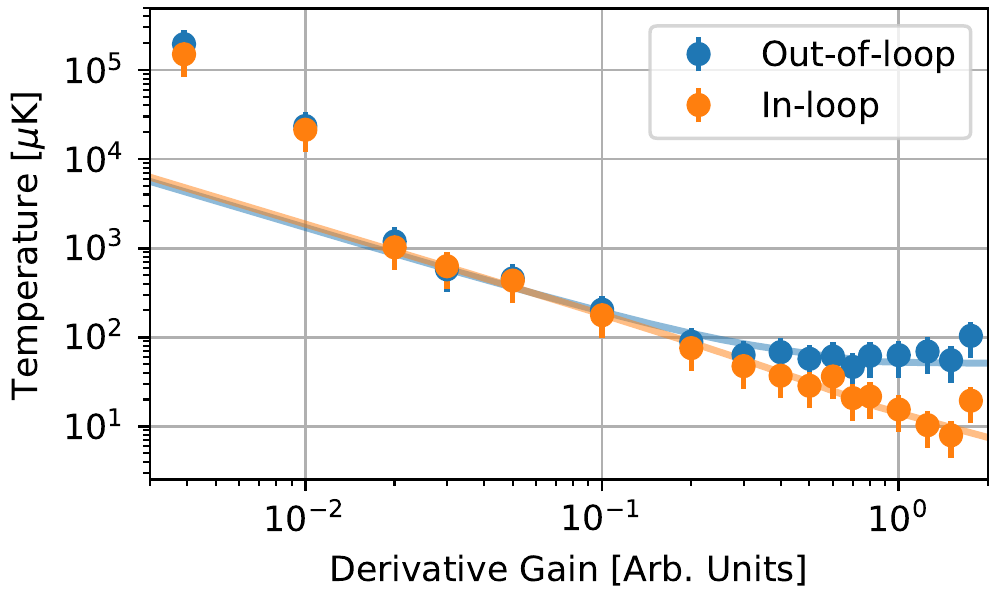}
    \caption{COM temperature measured in the $x$ direction for different feedback gains for the in-loop (orange) and out-of-loop (blue) sensors. The minimum temperature measured by the out-of-loop sensor is $50\pm 22\ \mu$K. The error is dominated by uncertainty in the mass of the sphere. The data corresponding to both sensors is well-described by a model of the feedback response at higher feedback gains. The temperatures measured at lower gains deviate from the expected model, possibly due to heating arising from digitization noise in the feedback system present at low gains.}
    \label{temp}
\end{figure}

Figure~\ref{psd} shows both the in-loop and out-of-loop amplitude spectral density of the sphere's displacement in the $x$ direction, $\sqrt{S_{xx}}$, as the gain of the feedback system is varied over several orders of magnitude. The COM temperature is obtained by first fitting each measurement of $\sqrt{S_{xx}}$ to the expected response of a harmonic oscillator driven by a thermal noise source with temperature $T$, $S_{xx} = \frac{2k_BT\Gamma}{MZ_0^2(\omega)}$~\cite{Flyvbjerg:2004,Li:2011}, where $k_B$ is Boltzmann's constant. Narrow noise features near the resonance peaks, including pickup at the line frequency of 60~Hz, are excluded from the fit. For the lowest feedback gains considered, broadening of the resonance feature is observed, preventing the spectrum from being fully described by the expression above. For the two data sets with the lowest feedback gains, e.g., the uppermost curve in Fig.~\ref{psd} and first two points in Fig.~\ref{temp}, the goodness-of-fit determined from the $\chi^2$ was poor for the fit to a single Lorentzian. For these, the convolution of a Lorentzian and a Gaussian broadening---a Voigt profile (with $\sigma \sim 10$ mHz)---was found to adequately describe the data. For all remaining feedback gains, where $\sigma \ll \Gamma$, the single harmonic oscillator form provided an acceptable fit. The best fit to the spectrum at each feedback gain determines the effective temperature $T$, shown in Fig.~\ref{temp}. Both the in-loop and out-of-loop sensors show a decrease in temperature as the magnitude of the feedback gain increases. At normalized feedback gains above $\sim 0.3$, the out-of-loop sensor no longer shows a continued decrease in the effective COM temperature, although the temperature measured by the in-loop sensor continues to decrease. Such decrease is due to unwanted cancellation of noise in the in-loop sensor by the feedback system known as ``noise squashing''~\cite{Rugar:2007,Quidant:2019,Novotny2019}. At these gains, the feedback system is unable to distinguish sphere motion from imaging noise and provides anti-correlated feedback that reduces the apparent noise in the in-loop sensor. Since the noise of the out-of-loop sensor is independent of the feedback system, it continues to provide an accurate measurement of the sphere even in the presence of squashing. The temperature measured from the out-of-loop sensor is minimum at $T = 50 \pm 22\ \mu$K, where the error is dominated by systematic uncertainty on the sphere's mass, arising from the possible variation in diameter and density~\cite{Acceleration_2017, Blakemore:2019}. Temperatures measured for normalized feedback gains below~$\sim 10^{-2}$ are higher than what is expected from the fit of Fig.~\ref{temp}. This is possibly caused by extra heating induced when the overall amplitude of the feedback approaches the digitization noise of the feedback system.

Improvements to the experimental setup have reduced the attainable temperature by more than an order of magnitude relative to the best previously demonstrated temperature for a sphere with diameter $> 1~\mu$m~\cite{Li:2011, Ranjit:2015}. The dominant remaining source of noise is found to be technical noise in the imaging beams, rather than damping from the residual pressure~\cite{beresnev:1990, Li:2011}, radiation pressure shot noise~\cite{Clerk:2010,Novotny2019,Novotny:2020}, or other fundamental noise sources. This can be observed in the lower plot of Fig.~\ref{psd}, where the signal measured by the in-loop photodiode without a trapped sphere (light blue data points) is similar to the signal measured with a sphere at high derivative gain. This noise may be improved with better isolation of the output optics after the vacuum chamber, by placing all the optics inside a sealed enclosure in the same style as the one used for the input optics before the vacuum chamber.

Despite this improvement in noise, the low resonance frequency of the trap leads to a substantial challenge in further reducing the effective temperature of such objects to near the ground state energy of the harmonic oscillator potential. While near ground state cooling~\cite{Novotny2019,Novotny:2020} and ground state cooling~\cite{Aspelmeyer:2019} have been recently observed for optically levitated nanospheres, it remains a daunting challenge to cool optically levitated microspheres to this regime. The COM temperature achieved here approximately corresponds to a mean occupation number $\langle n\rangle  \sim 1.6\times 10^4$. Although significant technical challenges must still be overcome, cooling of massive microspheres such as those considered here to the quantum regime may find application in tests of quantum superposition of massive objects and gravity~\cite{Anastopoulos:2015, Aspelmeyer:2018, Carlesso:2019}.

\section{Force and acceleration sensitivity}

The low trapping frequencies for which this system is optimized are advantageous for measurements of accelerations acting on the sphere at frequencies $\lesssim 100$~Hz. Such frequencies are typically in the range of interest for searching for new short-ranged interactions that couple to the mass or the number of nucleons in the sphere~\cite{Moore:2014,Rider:2016}. Previous work demonstrated that nanogram scale masses could be used to reach nano-$g$ acceleration sensitivities~\cite{Acceleration_2017}. Here, the upgraded system with reduced technical noise is used to reach pico-$g$ sensitivities in a $10^5$~s integration, where $g = 9.8$~m/s$^2$.

In addition to providing a measurement of the temperature of the sphere that is independent of the feedback, the out-of-loop sensor can also be used to improve the acceleration sensitivity when the system noise has a significant contribution from the imaging system rather than noise that affects the sphere motion directly. Since the imaging beam of the out-of-loop and in-loop sensors do not share the same optical path, they can be used to reject noise signals that are not common to both. This allows, for most frequencies in the band of interest between 1--100~Hz, an improvement of up to $\sim 10\%$ for the acceleration sensitivity relative to that obtained using a single sensor. 

\begin{figure}
    \centering
    \includegraphics[scale=0.85]{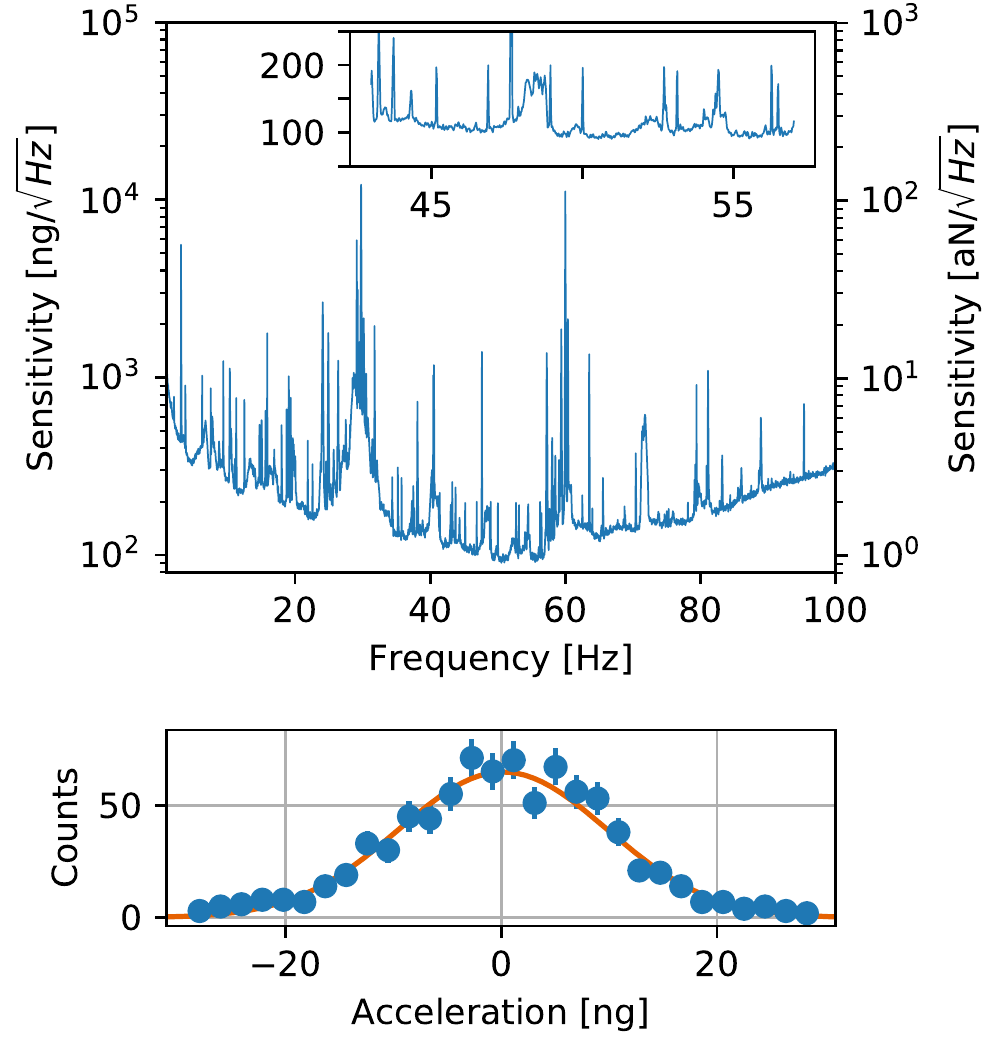}
    \caption{(Top): Acceleration (left vertical axis) and force sensitivity (right vertical axis) for a typical sphere.  The inset zooms in on the region near the minimum sensitivity at $\sim 50$~Hz. (Bottom): Distribution of the accelerations measured at 55~Hz during 52~s integration segments over a total of 12h in the absence of an applied force.}
    \label{acc}
\end{figure}

Figure~\ref{acc}~(top) shows the acceleration sensitivity spectral density (ASD) of a typical sphere. The ASD for this sphere is below $1~\mu g/\sqrt{\mathrm{Hz}}$ for frequencies spanning from 1--100~Hz and as low as $95\pm41$~n$g/\sqrt{\mathrm{Hz}}$ for frequencies from 40--60~Hz. This sensitivity represents an order-of-magnitude improvement over the current state of the art previously reported for spheres of this mass in Ref.~\cite{Acceleration_2017}. These values correspond to an overall force sensitivity of $0.95 \pm 0.11$~aN$/\sqrt{\mathrm{Hz}}$, comparable to the force sensitivity obtained for optically levitated nanospheres with mean diameters of $300$~nm~\cite{Ranjit:2016}. In addition to searches for new forces that scale with mass, this force sensitivity can enable searches for weak forces or impulses acting on massive objects, where large masses are needed to provide sufficiently large cross sections for rare interactions~\cite{carney:2019_1,carney:2019_2}.

The acceleration and force sensitivities measured near the resonance frequency (shown in the inset of Fig.~\ref{acc}~[top]) are consistent with $\sqrt{S_{FF}} = \sqrt{4k_B M T\Gamma} \sim 1$~aN/$\sqrt{\mbox Hz}$ for the temperature and effective damping $\Gamma$ measured in Figs.~\ref{psd} and~\ref{temp}. This agreement is consistent with the signal measured by the photodiode arising primarily from sphere motion for $\omega \sim \omega_0$. For frequencies farther from the resonance, the reduced signal amplitude leads to higher imaging noise.

Spheres in this setup can be reliably trapped for periods longer than a month at high vacuum, permitting long integration times for searching for periodic signals. Figure~\ref{acc}~(bottom) shows the distribution of the measured acceleration for repeated 52~s measurements over a period of 12 hours ($4.3 \times 10^4$~s) in the absence of an applied external acceleration and at a frequency of $f \sim 55$~Hz. A Gaussian fit results in a measured acceleration sensitivity of $170 \pm 340\ [stat] \pm 70\ [syst]$~p$g$, consistent with the measured ASD, and improving the sensitivity with the square root of the averaging time as expected for noise that is uncorrelated in time over multiple samples. The systematic error on the acceleration sensitivity arises mainly from the uncertainty in the sphere mass. In terms of force sensitivity, this corresponds $1.7\pm3.4\ [stat]\pm 0.2\ [syst]$~zN. The systematic error for the force sensitivity arises primarily due to uncertainties on the magnitude of the electric force applied during the calibration described above, and does not depend on the mass uncertainty.

\section{Trap stability}

\begin{figure}
    \centering
    \includegraphics[scale=0.85]{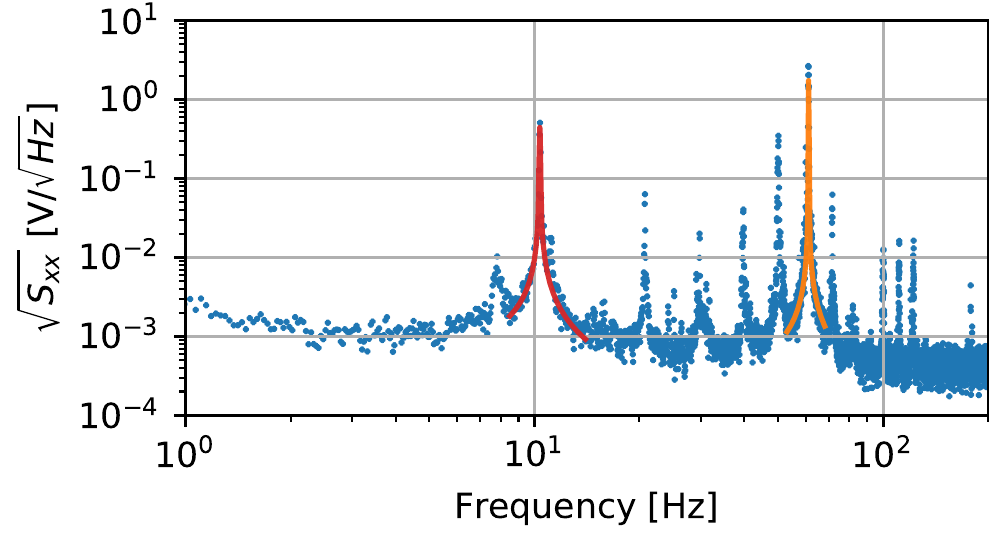}
    \caption{Power spectrum of the motion in the $x$ direction with no feedback cooling in all DOF. The calibration factor above does not apply as this sphere is free to levitate at any allowed position in the $z$ direction, resulting in a different optical gain. Several harmonics at $\omega/2\pi \sim 10$~Hz are observed as result of cross talk from the vertical motion of the sphere.}
    \label{nofb}
\end{figure}

The improved pointing stability demonstrated here also permits stable, high-vacuum optical trapping of microspheres for day-long periods without feedback cooling. Modern trapping experiments typically use commercially available SiO$_2$ spheres produced via the St\"ober process~\cite{STOBER:1968} that have substantially higher optical absorption than the high-purity materials used in the original demonstration of optical levitation in high-vacuum~\cite{Ashkin:1976}. In the present work, while feedback-free trapping of these microspheres is possible in high vacuum, active cooling is still required to maintain the sphere in the trap as it is pumped through the moderate vacuum pressures where photophoretic forces, sphere de-gassing~\cite{Acceleration_2017}, or other sources of noise not present in high vacuum may be significant. However, once the spheres are trapped at pressures $\sim 10^{-7}$~mbar, the feedback could be completely turned off in all DOF while maintaining stable trapping. Figure~\ref{nofb} shows the measured power spectrum of the motion of the sphere in the $x$ direction in the absence of any active cooling. To check the trap stability under these conditions, one of the spheres was kept in the trap for two days at $\sim 10^{-7}$~mbar without feedback. Following this test, the feedback system was re-engaged and the COM motion was again cooled to $\mu$K temperatures.

The above result is in sharp contrast to the behavior of the system before improvements were made to the trapping laser pointing stability, where the sphere was always lost $\sim$1~s after the active feedback was turned off at low pressures~\cite{Acceleration_2017}. Similar behavior has been reported in other recent experiments trapping nanospheres and microspheres in vacuum~\cite{Ranjit:2015,Ranjit:2016}. Since photophoretic forces are negligible in high vacuum, sphere loss is expected to be dominated by other mechanisms such as non-conservative forces~\cite{Roichman:2008, Ranjit:2016} and heating from technical noise in the trapping laser, among other sources. The results reported here are consistent with the dominant heating mechanism leading to loss of microspheres in previous work arising from laser pointing noise~\cite{Acceleration_2017}, since reduction of this heating mechanism permits stable trapping without feedback at high vacuum.

\section{Radiometric forces}
Current implementations of optical levitation in high-vacuum require feedback to damp the sphere's motion and maintain stable trapping. However, elimination of the need for such feedback could reduce the complexity of such systems, allowing for optical levitation in vacuum to be attained with a single, weakly focused laser and requiring no other optics or feedback electronics. Such low-complexity implementations may pave the way to large arrays of optically levitated microspheres~\cite{carney:2019_1,carney:2019_2} or more miniaturized apparatuses. Since current traps are loaded at pressures $\gtrsim 1$~mbar, the loss of spheres at intermediate vacuum pressures precludes a fully feedback-free system.

Initial studies of optical levitation in vacuum indicated that photophoretic forces may be responsible for losses upon pumping from atmospheric to vacuum pressures $\sim$1~mbar or lower~\cite{Ashkin:1971,Ashkin:1976, Ranjit:2015}. This force arise from absorption of the trapping laser light by the microsphere, leading to thermal gradients that produce photothermal forces on the sphere in the presence of background gas. Albeit small at atmospheric pressure, photophoretic forces can increase substantially as the pressure is reduced~\cite{Hettner:1926,Ashkin:1976}. 

\begin{figure}
    \centering
    \includegraphics[width=\columnwidth]{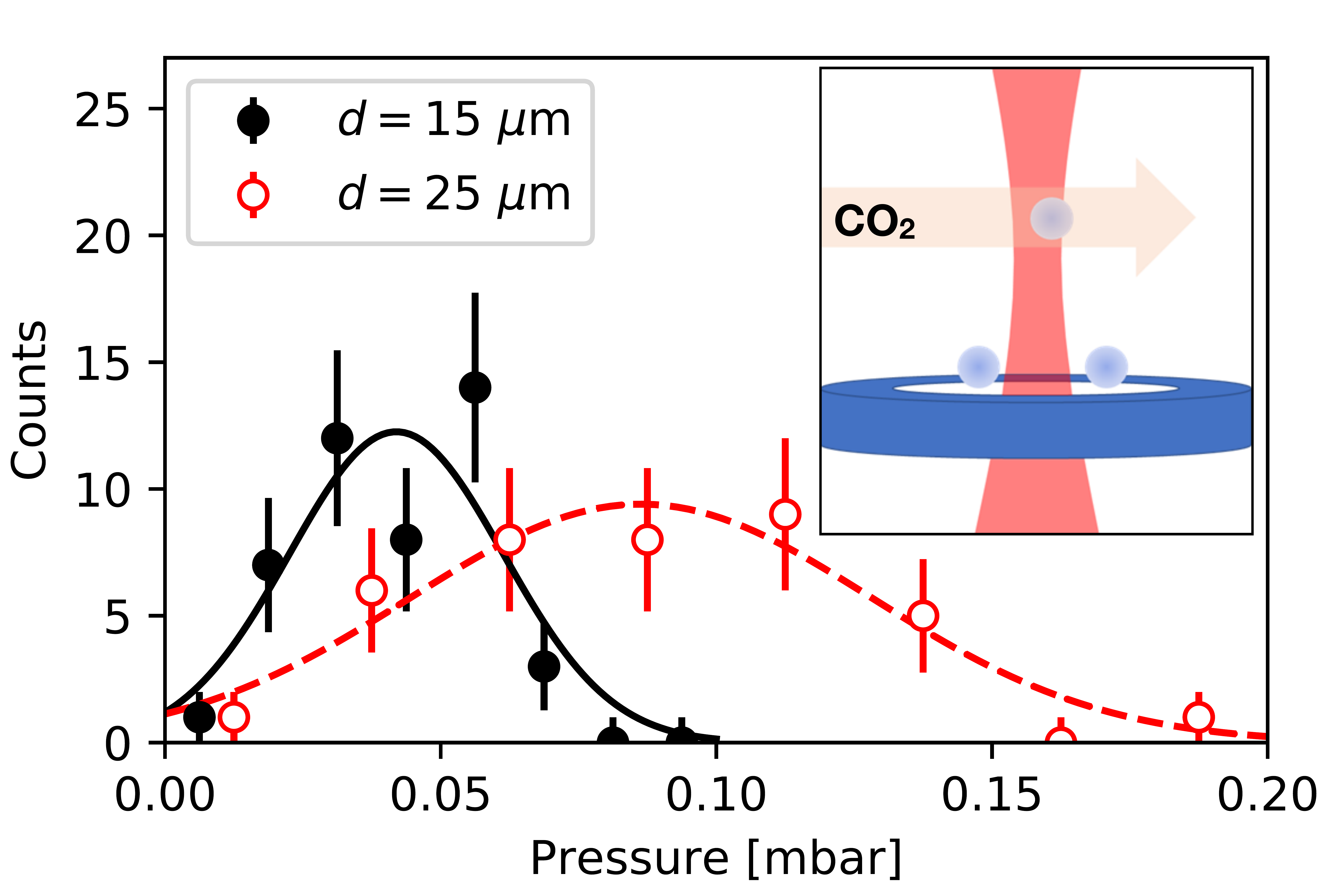}

    \caption{Distribution of pressures at which two spheres of different diameter are lost from the trap upon loading at atmospheric pressure and subsequent reduction in pressure until the sphere is lost (in the absence of additional heating from the CO$_2$ laser). The pressure distribution has a standard deviation of 0.02~mbar for the 15~$\mu$m sphere and 0.04~mbar for the 25~$\mu$m sphere. The inset shows a schematic of the setup used for repeated trapping of single spheres and the orientation of the CO$_2$ laser used for the studies of photophoretic forces.}
    \label{loss_press}
\end{figure}

An independent single-beam levitation trap~\cite{Ashkin:1971,LeBrun:2016} was developed to better characterize the interplay between photophoretic forces and laser noise and their impact on sphere loss. The two unique features of this setup, shown in the inset of Fig.~\ref{loss_press}, are: an additional CO$_2$ laser beam at 10.6~$\mu$m added to deterministically control the internal temperature and temperature gradients independently from the trapping beam power~\cite{Millen:2014}, and repeatable trapping of the same sphere by positioning it at a controlled location on a glass coverslip mounted just below the trap. With the use of optical stages, the coverslip could be positioned such that the same sphere was always in the path of the trapping beam when launched upwards into the trapping region by vibration of the coverslip with a piezoelectric actuator~\cite{LeBrun:2016}. This ability to repeatably trap the same sphere enables a study of the loss properties of the sphere from the trap that is free from sphere-to-sphere variations. While the loss pressure measured for different spheres from the same batch is found to vary over roughly an order of magnitude in pressure in our system, the variation in loss pressure for a single sphere that is repeatably trapped is substantially smaller. Figure~\ref{loss_press} shows a measurement of the pressure at which two different spheres (with diameters, $d \sim$15~$\mu$m and $d \sim25$~$\mu$m) are lost after loading into the trap at $\sim 1$~mbar and slowly reducing the pressure, while the CO$_2$ beam remains off. In each case, the same sphere was trapped $\sim$40--50 times. The 15~$\mu$m diameter sphere was lost from the trap between 0.02--0.07~mbar within 45 repetitions. The 25~$\mu$m sphere was typically lost at twice the pressure, with a similar fractional variation.

\begin{figure}[t!]
    \centering
    \includegraphics[width=\columnwidth]{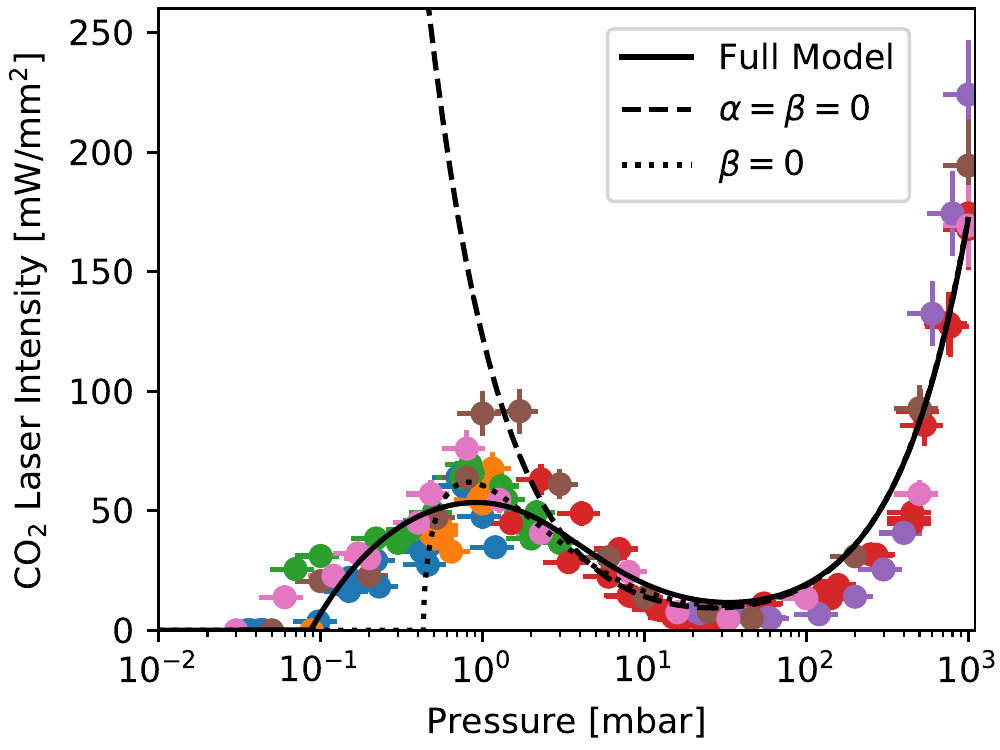}
    \caption{Measured CO$_2$ laser intensity needed for spheres to be lost from the trap as a function of pressure. The fits are based on phenomenological models for the heating of the COM temperature of the sphere, as described in the text. The full model considers photophoresis, heating from a pressure-independent term, and additional non-conservative forces arising from pressure-dependent displacements of the sphere. Different colors represent different spheres.}
    \label{co2}
\end{figure}

Several 15~$\mu$m diameter spheres were trapped in succession, and for each sphere the chamber pressure was varied between the loss pressure in the absence of the CO$_2$ laser and atmospheric pressure. At each pressure, the power of the CO$_2$ laser was increased until the sphere was lost from the trap. Upon loss from the trap, the same sphere was again loaded into the trap and the measurement was repeated at a different pressure. Figure~\ref{co2} shows the CO$_2$ laser intensity measured upon loss of the sphere at a variety of pressures and for several spheres with diameter of 15~$\mu$m. For these measurements, the maximum CO$_2$ laser intensity never exceeds 250~mW/mm$^2$. The maximum total power absorbed from the CO$_2$ laser is estimated to be $\sim$30~$\mu$W~\cite{Yang:10}, which is comparable to the power absorbed from the 980~nm trapping beam~\cite{Acceleration_2017}. Despite similar absorbed powers, the absorption per unit area is approximately uniform along the sphere for the trapping beam geometry as a consequence of the absorption coefficient $\alpha \ll d^{-1}$ at 980~nm and also due to the low NA of the trapping beam. In contrast, most of the power from the CO$_2$ laser is absorbed in the first half of the sphere since $\alpha > d^{-1}$ at 10.6~$\mu$m. In the former case, the symmetric absorption of the 980~nm trapping light results in a uniform temperature distribution on the surface of the sphere.  In the latter case, the asymmetric absorption of the weak 10.6~$\mu$m light produces a significant gradient in the surface temperature in the direction of the incident beam, which is ideal for studying the forces that couple to such temperature gradients~\cite{Hettner:1926}. In addition, due to the high absorption and low power of the CO$_2$ laser, the maximum force it exerts due to radiation pressure is $\sim$0.1~pN, which is at least an order of magnitude smaller than the optical forces from the trapping laser.

To estimate the overall sphere temperature and gradient caused by the CO$_2$ laser and trapping beams, finite element method simulations were performed using COMSOL. The simulations model the heating within the volume of the sphere from absorption of both lasers, as well as heat transfer from the sphere surface to the surrounding air under the assumption that the sphere temperature has reached equilibrium. At high pressures ($\sim$1~atm), the COMSOL simulation models the convective and conductive heat transport from the sphere. At lower pressures, the thermal conductivity of the residual air is estimated following~\cite{Ganta:2011}.

 These simulations show a roughly pressure independent internal temperature gradient that is directly proportional to the CO$_2$ laser intensity. These simulations agree with the assumptions originally employed to estimate the radiometric forces on optically levitated objects~\cite{Ashkin:1976} and disagree with more recent analyses where a pressure-dependent thermal gradient is assumed~\cite{Ranjit:2015}. The pressure-independent gradient is observed even for lower pressures where the sphere temperature increases above that of the background gas. This behaviour results from the approximately spherically symmetric transfer of heat to the background gas and to thermal radiation in the steady state. This spherically symmetric heat transfer is valid whenever the temperature gradients are small compared to the overall temperature, as is typical for optically levitated objects. In this case, the gradients are determined only by the spatial profile of laser absorption and thermal conductivity within the sphere itself. In this model, the photophoretic force is proportional to the CO$_2$ laser intensity and has a pressure dependence given by~\cite{Hettner:1926,Ashkin:1976}:

\begin{equation}
    F = \pi\eta r\sqrt{\frac{\gamma k_B}{4 m_g T_g}}\Delta T\left(\frac{p_0 p}{p_0^2 + p^2}\right) \propto I_{CO_2}\left(\frac{p_0 p}{p_0^2 + p^2}\right),
\end{equation}

\noindent where $\Delta T$ is the temperature difference across the sphere, $I_{CO_2}$ is the CO$_2$ laser intensity, $m_g$ is the molecular mass of the gas, $\gamma$ is the accommodation factor, $\eta$ is the gas viscosity, $T_g$ is the gas temperature, $r$ is the sphere radius and $p_0 = \frac{3\eta}{r}\sqrt{\frac{k_B T_g}{m_g\gamma}}$.

In order for the sphere to be lost from the trap, the work done by the photophoretic force and the total initial COM energy (resulting from thermal motion due to the background gas and any other sources of heating) must together overcome the depth of the optical potential $E_{trap}$. A phenomenological model that describes the data shown in Fig.~\ref{co2} can therefore be written:

\begin{equation}
\begin{split}
    (F+F_{abs})\Delta x &\geq E_{trap} - \frac{\Gamma k_B T_{g} + \alpha k_B T_{heat} + \beta \Delta x}{\Gamma + \alpha}, \\
    & \Delta x \sim \frac{F+F_{abs}}{M\omega_0^2}
\end{split}
\end{equation}
where $F_{abs}$ is the force due to the radiation pressure from the absorption of the CO$_2$ beam, $\Gamma$ is the pressure-dependent damping rate due to the residual gas~\cite{beresnev:1990,Li:2011}, $T_{heat}$ is the effective temperature of a pressure independent source of COM heating with damping $\alpha$ (e.g., heating due to technical noise in the laser). However, these terms alone cannot fully account for the data, in particular in the low pressure regime just above where the spheres are lost in the absence of the CO$_2$ laser. The addition of a parameter, $\beta$, which describes the rate of work done by the non-conservative optical scattering force when the sphere is displaced by an amount $\Delta x$ from the trap equilibrium due to the photophoretic force from the CO$_2$ laser~\cite{Roichman:2008, Ranjit:2016} provides an acceptable fit to the data over the full range of the measurement (Fig.~\ref{co2}).

Considering only photophoretic forces ($\alpha = \beta = 0$), this model can fully describe the data at pressures $\gtrsim$~1~mbar, where the model predicts a maximum for the photophoretic force at $p_0 \sim$ 30~mbar. The fit in this region finds $\gamma = 0.66\pm0.08$ which is within the range of accommodation coefficients previously measured for SiO$_2$~\cite{Ganta:2011}. The photophoresis-only model also predicts that the intensity of the CO$_2$ laser required to lose the sphere from the trap should increase for pressures below $\sim$1~mbar as the gas becomes more rarefied and photophoretic forces are reduced, in disagreement with the data.

Adding an additional pressure-independent heating term to the model ($\alpha\neq0$, $\beta=0$) can reproduce the finding that the CO$_2$ loss intensity goes to zero at low pressure, as long as the energy provided by the external source is larger than the trap depth, i.e, for low enough pressures where $\Gamma T_{g} \ll \alpha T_{heat}$ and for $T_{heat} > E_{trap}/k_B$. The data indicates that this occurs at pressures $\lesssim$ 0.1~mbar.  Although the best-fit model with the inclusion of this term agrees qualitatively with the data, it cannot account for the quantitative behavior of the CO$_2$ laser loss intensity in the region $\lesssim 1$~mbar since it predicts a steeper pressure dependence than observed. Similar pressure dependence can also be obtained by considering a heating rate that is proportional to the internal temperature of the microsphere.  While simulations indicate that the overall microsphere temperature can begin to rise at these pressures due to the absorption of the trapping beam~\cite{Millen:2014,Novotny_intT:2018}, a simple model for this heating that is proportional to the mean sphere temperature (rather than the surface temperature gradient) fails to reproduce the measured data, implying another mechanism for the measured pressure dependence.

The full model shown in Fig.~\ref{co2}, also considers a non-negligible contribution from heating that is proportional to the displacement of the sphere COM due to the photophoretic force induced by the CO$_2$ laser. With the inclusion of this additional parameter, $\beta$, a fit to the data can provide reasonable agreement over the full range of pressures. Such forces could arise from work done on the COM motion of the sphere by the scattering force from the trapping laser during its motion as it is displaced in the direction perpendicular to the trap by photophoretic forces~\cite{Roichman:2008, Ranjit:2016}.  Heating from such non-conservative scattering forces has been observed in fluid-based traps~\cite{Roichman:2008}, and the effect of photophoretic forces could be enhanced at moderate vacuum pressures through such a mechanism. 

The model above predicts that spheres are lost from the trap at sufficiently low pressures if $T_{heat} > E_{trap}/k_B$. The same model is also consistent with stable trapping of microspheres in high vacuum without feedback cooling as long as $T_{heat} < E_{trap}/k_B$. The trap depth, $E_{trap}$, can be significantly larger for the microspheres used in this work than for smaller optically levitated objects, allowing this condition to be more easily met.

For the setup presented in the first part of this work, spheres can be stably trapped at pressures around $p_0$ and at high vacuum even in the absence of feedback cooling but are typically lost at intermediate pressures if the feedback is disabled. This observation indicates that photophoretic forces are not dominant at any pressure, consistent with the negligible thermal gradients simulated from absorption of the low NA trapping beam. The observation that spheres can be reliably trapped in high vacuum without feedback cooling further indicates that pressure-independent heating sources such as the pointing noise of the trapping beam are also not the cause for the loss of the spheres at intermediate pressures. An alternative possible mechanism for loss during pump down with no feedback may be associated with sphere degassing due to the high water content~\cite{STOBER:1968} and high internal temperature at these pressures~\cite{Acceleration_2017}. This possibility for sphere loss (although not directly arising from photophoretic forces) remains consistent with previous results~\cite{Ashkin:1976} in which spheres made of high-purity, low-absorption materials were successfully pumped through intermediate pressures in the absence of feedback cooling. Further work to develop nanogram scale objects from such high purity materials may result in a better understanding of the loss mechanism for this kind of optical trap and could enable feedback free optical trapping over all ranges of vacuum pressures.

\section{Conclusion}
This work reports acceleration and force sensitivity for optically levitated nanogram spheres that reach the pico-$g$ and zN scale, representing more than an order-of-magnitude improvement over past measurements for optically levitated micron-sized objects~\cite{Ranjit:2015,Acceleration_2017, Rider:2017}. An effective COM temperature of $T = 50 \pm 22\ \mu$K is reached, which is more than an order of magnitude lower than previously reported temperatures for objects $>1\ \mu$m in diameter~\cite{Li:2011,Li:2013}. The high sensitivity and low temperatures are a consequence of improved stability of the trapping laser and low system noise. $\mu$K temperatures are obtained even for electrically neutral objects, in contrast to other demonstrations that use electric feedback forces and require the spheres to have a net charge~\cite{Quidant:2019, Novotny2019,Aikawa:2019}. Cooling of electrically neutral spheres can minimize coupling to stray electric fields and other sources of noise that are typically present in sensing applications that use optically levitated objects. 

Substantial further advances are required to cool nanogram-scale objects to temperatures near their ground state. Such improvements would also be expected to improve the overall force and acceleration sensitivity. Since the sensitivity obtained here is found to be limited by technical noise in the imaging system, the first step towards further improvement includes the reduction of the pointing noise from the beams used to image the sphere. This increased sensitivity will improve precision tests of the electrical neutrality of matter~\cite{Morpurgo:1977, Morpurgo:1984, Baumann:1988,Bressi:2011}, searches for millicharged dark matter particles bound in the matter~\cite{HOLDOM:1986,Moore:2014}, and tests of Newton's and Coulomb's laws at micron distances~\cite{Nelson:2003, Geraci:2008, Rider:2016, Deca:2016, Yoshioka:2018}.

The low noise of the system also allows for stable, feedback-free trapping of microspheres in high vacuum.  This observation reproduces the feedback-free trapping demonstrated in the pioneering investigation of optical levitation in high vacuum by Ashkin and Dziedzic~\cite{Ashkin:1976} and further extends feedback-free levitation to higher absorption materials and to stable trapping of nanogram objects for time periods exceeding a day. Feedback-free levitation may be useful for applications that require high sensitivity, but do not require the effective COM temperature to be substantially cooled, such as detection of particle recoils in large arrays of levitated spheres~\cite{carney:2019_1,carney:2019_2} or micron-scale pressure gauges~\cite{blakemore_gauge:2019}. In addition, the elimination of the requirement for active feedback may enable simple implementations of optical trapping in high vacuum relevant for inexpensive, miniaturized force sensors or accelerometers.

Despite feedback-free, stable levitation at high and low vacuum, spheres are lost from the trap while at intermediate pressure. Photopheretic forces arising from internal temperature gradients are shown to have maximum effect at $\sim$30~mbar and to vanish at smaller pressures, but are found to be too small to account for sphere loss in the intermediate pressure range for the setups described here. These results indicate that photophoresis and laser technical noise are unlikely to be the dominant loss mechanisms for these traps, and other effects are expected to account for the sphere loss in this regime.

\section{acknowledgments}
We would like to thank the Gratta group at Stanford for useful discussions related to this paper. This work is supported, in part, by the Heising-Simons Foundation, NSF Grant No. 1653232, the Alfred P. Sloan Foundation, and ONR Grant No. N00014-18-1-2409.

\bibliographystyle{apsrev4-1}
\bibliography{references_new}{}

\end{document}